 \documentclass[final,3p,times,twocolumn,authoryear]{elsarticle}

\usepackage{graphicx}
\usepackage{amsmath}
\usepackage{textcomp}
\usepackage{amssymb}
\usepackage{lineno}
\usepackage[version=3]{mhchem} 

\journal{Earth and Planetary Science Letters}

\begin{document}

\begin{frontmatter}


\title{Atmospheric nitrogen evolution on Earth and Venus}
\author{R.~D.~Wordsworth}
\ead{rwordsworth@seas.harvard.edu}
\cortext[cor1]{Corresponding author.}
\address{Harvard Paulson School of Engineering and Applied Sciences, Harvard University, Cambridge, MA 02138, \\  Department of Earth and Planetary Sciences, Harvard University, Cambridge, MA 02138}
\author{}
\address{}

\begin{abstract}
Nitrogen is the most common element in Earth's atmosphere and also appears to be present in significant amounts in the mantle. However, its long-term cycling between these two reservoirs remains poorly understood. Here a range of biotic and abiotic mechanisms are evaluated that could have caused nitrogen exchange between Earth's surface and interior over time. In the Archean, biological nitrogen fixation was likely strongly limited by nutrient and/or electron acceptor constraints. Abiotic fixation of dinitrogen becomes efficient in strongly reducing atmospheres, but only once temperatures exceed around 1000~K. Hence if atmospheric \ce{N2} levels really were as low as they are today 3.0 -- 3.5~Ga, the bulk of Earth's mantle nitrogen must have been emplaced in the Hadean, {most likely at a time when the surface was molten}. The elevated atmospheric \ce{N} content on Venus compared to Earth can be explained abiotically by a water loss redox pump mechanism, where oxygen liberated from \ce{H2O} photolysis and subsequent H loss to space oxidises the mantle, causing enhanced outgassing of nitrogen. This mechanism has implications for understanding the partitioning of other Venusian volatiles and atmospheric evolution on exoplanets.
\end{abstract}


\end{frontmatter}


Despite the fact that it makes up over 78\% of our atmosphere by volume and is essential to all known life, nitrogen remains a poorly understood element on Earth. In comparison to other major non-metal elements such as oxygen and carbon, the mechanisms responsible for its initial delivery, isotopic evolution and partitioning between the surface and mantle are all still subject to great uncertainty. As a result, one of the most basic features of our planet's climate -- the atmospheric pressure -- remains unexplained by Earth science.

Nitrogen is an important player in climate because it causes pressure broadening of the absorption lines of greenhouse gases like \ce{CO2} and \ce{H2O} \citep{Goldblatt2009}, and can also cause intense warming in combination with \ce{H2} in reducing atmospheres via collision-induced absorption in the 800-1200~cm$^{-1}$ \ce{H2O} `window' region of the infrared spectrum \citep{Wordsworth2013a}. The atmospheric nitrogen inventory also affects the latitudinal temperature gradient (in general denser atmospheres transport heat from equator to poles more effectively). All these effects are particularly relevant in the context of the faint young Sun problem, which arises because solar luminosity was 20-25\% lower 3-4~Ga, but Earth was not permanently glaciated in the Archean \citep{Saga:72}. Furthermore, planets with very low atmospheric \ce{N2} will no longer cold trap \ce{H2O} and hence will irreversibly oxidize via hydrogen loss, which can lead to the buildup of abiotic \ce{O2}-dominated atmospheres in extreme cases \citep{Wordsworth2014}. Understanding Earth's nitrogen is hence important both as a fundamental problem, and for addressing wider questions of planetary climate evolution.

Thanks to its strong triple bond and lack of a permanent dipole moment, molecular nitrogen (\ce{N2}) is both chemically unreactive and highly volatile, and hence was once thought of as similar to the noble gases in terms of its incompatibility in the solid Earth. However, this simple view of \ce{N2} has been eroded over the last few decades by {a number of studies suggesting that} a significant fraction of the present-day atmospheric inventory {(between around 0.4 and 7 times)} is currently stored in the crust and mantle \citep{Marty1995,Marty2003,Halliday2013,Johnson2015}. The main evidence for this comes from the \ce{N2}/\ce{^{40}Ar} ratios measured in mid-ocean ridge basalts (MORBs) and rocks of mantle plume origin, which allow the mantle N inventory to be estimated when combined with bulk silicate Earth (BSE) estimates of radiogenic \ce{K}  abundance. 
 {Earth's atmosphere is also depleted in \ce{^{14}N}  relative to chondritic C and noble gas abundance ratios \citep{Marty2012} (the `missing N' problem), which can be explained by the presence of a  substantial mantle N component \citep{Johnson2015}.}
The correlation of N with \ce{^{40}Ar} in MORBs and the large-ion lithophile elements in metamorphic rocks also suggests that its primary form in the crust and upper mantle is \ce{NH4+}, where it generally substitutes for \ce{K+}  \citep{Busigny2013}. 
Finally, measurements of the content of nitrogen and other volatiles in Alpine metasediments suggest that today, there is a net flux of \ce{N} to the mantle of order $1\times 10^{12}$~g/yr \citep{Busigny2013}. 

Based on these observations, it has been proposed that biological N fixation has caused atmospheric nitrogen levels to decrease over geological time, and hence that surface pressure on the early Earth was two times or more the present-day value \citep{Goldblatt2009}. However, this apparently logical conclusion has not been supported by recent observational constraints on the ancient atmospheric pressure. First, measurements of the radii of putative fossil raindrops have been argued to constrain the atmospheric density 2.7~Ga to no more than double the present-day value \citep{Som2012}. Independently, recent measurements of \ce{N2} to \ce{^{36}Ar} ratios in the fluid inclusions in Archean hydrothermal quartz have suggested that as long ago as 3.0-3.5~Ga, the partial pressure of \ce{N2} in the atmosphere was 0.5-1.1~bar \citep{Marty2013}. Hence Earth's mantle N inventory was apparently in place by the mid-Archean or earlier. If these measurements are correct, nitrogen exchange between the mantle and interior must have been very effective at some {earlier} point in Earth's early history.

Further clues to terrestrial nitrogen's origins and evolution come from isotopic measurements. Of nitrogen's two stable isotopes, the rarer \ce{^{15}N} is enhanced in the atmosphere and in crustal rocks relative to the mantle [\ce{\delta^{15}N} of approx. $-5$\textperthousand$ $ vs. the atmosphere in the upper mantle and perhaps as low as $-40$\textperthousand$ $ in the lower mantle; \cite{Cartigny2013}]\footnote{{Here for a given sample $i$, $\ce{\delta^{15}N}=1000[(^{15}\ce{N}/^{14}\ce{N})_i / (^{15}\ce{N}/^{14}\ce{N})_{atm}-1]$ and $(^{15}\ce{N}/^{14}\ce{N})_{atm} = 3.7\times10^{-3}$ is the Earth atmospheric value.}}. 
This has been interpreted previously as an indication of biological processes.  Earth's atmospheric \ce{^{15}N}/\ce{^{14}N} ratio ($\sim 3.7\times10^{-3}$) is close to the values found in ordinary and carbonaceous chondrites \citep{Marty2012}, but the solar and gas giant N reservoirs are lighter \citep[$\sim2.3\times10^{-3}$; ][]{Owen2001}. This has led to the prevailing view that most of Earth's \ce{N} was delivered by bodies with composition close to that of the carbonaceous chondrites. At which stage of accretion, and in what form this nitrogen was delivered is still uncertain, but a reasonable interpretation is of volatile-rich bodies {of chondritic composition} impacting the Earth during the later stages of accretion, with much of the nitrogen in the form of ammonia ices or simple organic compounds such as HCN. 

Clearly, our current understanding of the long-term atmosphere-interior exchange of nitrogen on Earth is lacking in many respects. Most previous research has focused on improving the (vital) geochemical constraints on the deep-time N cycle. To date, however, far less attention has been paid to the theoretical aspects of the problem. Here the importance of several key processes to Earth's earliest nitrogen cycle are analyzed (see Fig.~\ref{fig:photo_schem}). First, it is noted that nitrogen delivered to Earth in fixed form would have initially degassed into the atmosphere and that it therefore must have entered the mantle at some point after this. 
Based on primary productivity constraints and other arguments, it is then argued that drawdown of bars of nitrogen before 3.0-3.5~Ga via biological fixation is highly unlikely. An atmospheric chemistry model is used to show that abiotic reduction of atmospheric nitrogen can be effective, but only under conditions where the atmosphere is strongly reducing and the surface temperature is extremely hot. Based on this, a new explanation involving a magma ocean redox pump driven by water loss is proposed for the differing atmospheric N inventories of Earth and Venus.

\section{Nitrogen delivery by impacts}\label{subsec:impacts}

{The first stage of Earth's nitrogen cycle is delivery of the element during accretion.} As discussed in the introduction, most of this incoming nitrogen was likely fixed in the form of ammonia ices or organics. However, the temperatures and peak shock pressures on impact for the majority of accreting bodies mean that it would have initially thermalized and degassed into the atmosphere. 
{Impact shock experiments on carbonaceous chondrites \citep{Tyburczy1986} indicate that for accretion occurring after Earth has reached around 30\% of its final radius, mineral devolatilization is effective. As the majority of Earth's nitrogen was probably delivered late, this implies } the immediate post-impact location of delivered nitrogen would have been the atmosphere. {Note that as well as depositing their own nitrogen to the atmosphere, impactors would also devolatilize any crustal and sedimentary nitrogen that was already present in the region of impact. We do not attempt to model this process here, but it would act to reduce the efficiency of the slow drawdown mechanisms discussed in the next two sections.}

\section{Direct biological nitrogen fixation} 

{If most or all of the N currently in the bulk silicate Earth started in the atmosphere, some mechanism must be responsible for depositing it in the mantle over time.} The simplest explanation for the origin of Earth's mantle N is the same process that is apparently producing a net subduction flux of nitrogen today: biological N fixation. Is this likely to have been important throughout geologic time? If the present-day flux were constant throughout Earth's history, 22-35\% 
of the present-day atmospheric inventory would have been drawn down between the cooling of Earth's surface after the Moon-forming impact 4.4~Ga (the earliest possible date for the origin of life) and 3.0-3.5~Ga, when recent studies \citep{Marty2013}  argue for close to present-day atmospheric levels of \ce{N2}. Clearly, higher subduction fluxes in the past are required to explain the current mantle \ce{N} inventory.

Several effects could have altered the biogenic flux of fixed N to subduction zones with time.  In the oceans today, the molar ratio of N to C in organic matter is approximately 1:6.6 (the Redfield ratio). In the past, this ratio might have been slightly different due to e.g. variations in the contribution of protein to the total biomass \citep{Geider2002}. However, it is highly unlikely to have decreased below around 1:4, due to fundamental amino acid stochiometry. 

Changes in the rates of sediment nitrogen loss prior to subduction are harder to constrain. Under anoxic conditions, conversion of fixed N to nitrate and hence denitrification rates would clearly have been less efficient than today. However, ammonification of organic nitrogen would still have provided the biosphere with an efficient mechanism to return reduced N to the ocean and atmosphere. Furthermore, fixed N is energetically costly, so an early biosphere with a much greater organic N:C loss ratio than today would contain strong incentives for the development of efficient sediment N removal pathways if they did not exist initially.

If the N:C ratio in sediment did not change significantly in the past, what about the total rate of subduction of organics? Simple interpretations of the isotopic record of sedimentary carbon imply that organic carbon burial rates were around 10\% of the total in the Archean, compared to around 20\% today \citep{DesMarais1992}. Seafloor carbonization and authigenic carbonate formation \citep{Bjerrum2004,Schrag2013}, which are both poorly constrained, would decrease this percentage further if they were significant in Earth's early history. 

Lower Archean organic carbon burial rates make sense even if oxygenic photosynthesis emerged early, because nutrient supply to the biosphere in the Archean was probably lower than today. The present-day net primary productivity (NPP) of the marine biosphere is limited on long timescales by the supply of phosphorous \citep{Filippelli2008}. The dependence of phosphorous supply on land weathering means that the area of exposed land is a key factor controlling primary productivity.  In the Archean the continents were still forming, with estimates of the amount of emergent land 3.5~Ga ranging between 3 and 30\% \citep{Flament2008,Dhuime2012}. This implies decreased P supply to the oceans \citep{Wordsworth2013a}. In addition, abiotic sinks for \ce{P} are probably higher in anoxic environments \citep{Bjerrum2002}.

Perhaps the most important point regarding biological N fixation in the Archean, however, is that today's biosphere is powered by oxygenic photosynthesis. The date when this key metabolism emerged is not known, but there is no strong evidence that it existed 3.0~Ga or earlier. This is crucial, because before the advent of oxygenic photosynthesis primary productivity must have been drastically lower. Ecological models \citep{Kharecha2005} predict that in an ecosystem dominated by methanogenesis, NPP was lower than today by a factor of 500-3000. Under these circumstances, burial rates of biologically fixed N would have to be significantly lower as well. Substantial nitrogen subduction due to direct biological fixation in the early Archean and Hadean therefore appears unlikely.

\section{Photochemical nitrogen fixation}

If methanogens emerged prior to oxygenic photosynths, there was probably a period when atmospheric methane levels were high \citep{Pavlov2003}. Then, HCN formation via reactions of the type
\begin{equation}
\ce{^3CH_2 + N(^4S) \to HCN + H} \label{eq:HCN} 
\end{equation}
in the upper atmosphere could potentially have led to rapid N fixation. {Reaction (\ref{eq:HCN}), which was first studied in a planetary atmospheric context by \cite{Yung1984}, has been proposed as an important source of HCN for prebiotic chemistry in Earth's early atmosphere \citep{Zahnle1986}. Given sufficient quantities of atmospheric methane, it could have led to N fixation rates of up to $1\times10^{10}$~molecules/cm$^2$/s \citep{Tian2011}. In the absence of any recycling, this would be sufficient to remove Earth's entire present-day nitrogen inventory in around 100~Myr \citep{Tian2011,Wordsworth2013a}.}

In the context of bulk N drawdown, the advantage of atmospheric HCN formation over surface biological fixation of N is that the rate of methane production in a biosphere dominated by methanogens and/or anoxygenic photosynthesis is many times greater than the NPP \citep{Kharecha2005}. Could this provide an alternative route to drawdown of the nitrogen in the early atmosphere? 

To work before 3.0-3.5 Ga, this mechanism clearly requires either methanogens or anoxygenic phototrophs (or both) to have created a globally significant biosphere very early on. This is possible, as methanogenesis is believed to be an ancient metabolism, although phyllogenetic molecular clock estimates of the date of mesophilic methanogen emergence range from 3.8 to 1.3 Ga \citep{Battistuzzi2004,Blank2009}.  
The high \ce{HCN} production rates needed to achieve rapid N drawdown also require extremely high atmospheric methane levels ($\sim$1000~ppmv). 
Besides a thriving methanogen ecosystem, this requires a combination of high hydrogen outgassing and seafloor serpentinization rates and/or low hydrogen escape rates. 

Finally, once it reaches the surface, the photochemically produced \ce{HCN} must still be incorporated into sediment. The most logical way for this to occur is by biological conversion to ammonia [e.g., as occurs via certain fungi and bacteria today; \citep{Knowles1988}], followed by incorporation into organic matter. However, the biosphere would be unlikely to incorporate nitrogen into organic material at a rate greater than that required by the NPP, which in a fixed-N-rich environment would probably be P-limited. Then, the restrictions on N burial based on availability of nutrients and/or electron donors discussed in the last section would again apply. Even if the biosphere were able to incorporate all the available nitrogen from \ce{HCN}, the conversion efficiency from \ce{HCN} to \ce{NH3} to organic matter would need to be high, because gaseous ammonia is highly unstable to dissociation by photolysis in the atmosphere {next section. Hence rapid drawdown of N before 3.0-3.5~Ga by this mechanism also appears challenging to achieve. For this reason, we now turn to abiotic processes that could have been important in the Hadean. }

\section{\ce{N2} thermolysis in a highly reducing atmosphere}\label{subsec:NH3make}

In the Hadean, atmospheric and crustal conditions would have been far more reducing than later on, due to processes such as nebular capture,  \citep{Genda2008}, 
chemical equilibration of accreting chondritic material,  \citep{Schaefer2010}, 
 or serpenitization. Then, reduced forms of nitrogen such as \ce{NH3} would be thermochemically favoured near the surface.  Once reduced, nitrogen could have become incorporated into ocean sediment and subducted, or dissolved directly into the mantle, {if the surface was hot enough to be in a magma ocean state.}

To investigate this scenario quantitatively, it is necessary to work out the conditions under which near-surface processes could cause rapid reduction of \ce{N2}. Reduced nitrogen in the form of ammonia (\ce{NH3}) is highly vulnerable to photolysis by UV radiation in the upper atmosphere \citep{Kuhn1979}, 
so given low thermochemical conversion rates even a hydrogen-rich atmosphere would have very low surface abundances of reduced nitrogen. To assess the conditions under which equilibrium chemistry would dominate in the lowest part of a highly reducing atmosphere, {we first calculate the rate of \ce{NH3} photolysis in the high atmosphere, followed by the rates of thermochemical N reduction as a function of temperature. {The question of how reduced atmospheric N is transferred to the mantle is tackled afterwards: the answer turns out to be strongly constrained by the temperatures at which surface reduction of N is effective.}

\subsection{Destruction of reduced nitrogen by photolysis}

An upper bound on the maximum possible global average loss rate of \ce{NH3} from photolysis can be derived from the expression
\begin{equation}
\Phi_{photo,max}= \frac 14 \int_0^{\lambda_{cut}}Q_y(\lambda)\mathcal F_\odot(\lambda)\mbox{d}\lambda,
\label{eq:photolossmax}
\end{equation}
where $\mathcal F_\odot(\lambda)$ is the early solar UV flux as a function of wavelength $\lambda$, $Q_y(\lambda)$ is the quantum yield of reaction \mbox{\ce{NH3 + h\nu \to NH2 + H}},  $\lambda_{cut}$ is the wavelength above which the photolysis absorption cross-section of \ce{NH3} drops to negligible values, and the factor $1/4$ accounts for averaging over the sphere. Using the best estimate for the XUV/UV spectrum of the Hadean Sun (see below), assumed quantum yield of 1.0 and $\lambda_{cut}=230$~nm, equation (\ref{eq:photolossmax}) yields $3.8\times10^{13}$~molecules/cm$^2$/s. 

To {investigate Hadean \ce{NH3} photolysis in more detail}, a photochemical code was constructed. The code solves the 1D partial differential equation
\begin{equation}
\frac{\partial n_i}{\partial t} = P_i - L_i n_i - \frac{\partial \Phi_i}{\partial z},
\label{eq:photochem1}
\end{equation}
\begin{equation}
\Phi_i = -\kappa_D n  \frac{df_i}{dz} 
\label{eq:photochem2}
\end{equation}
in time on a 100-layer vertical grid that extends from the surface to a height $z_{top}$, defined here as the altitude where the atmospheric pressure $p$ is 1~Pa. In (\ref{eq:photochem1}-\ref{eq:photochem2}), $n$ and $n_i$ are the total number density and number density of a given species, respectively, in molecules~cm$^{-3}$, $f_i$ is the molar concentration of a given species, $P_i$ and $L_i$ are chemical production and loss terms, respectively, and $\kappa_D$ is the (eddy) diffusion coefficient. $\kappa_D$ was taken to be a constant $1\times10^5$~cm$^2$~s$^{-1}$ in some simulations, while in others the present-day $\kappa_D$ profile from \cite{Hunten1975} was used. Each simulation was run until a steady state was reached ($\partial_t n_i=0$).

In the model, surface pressure $p_s$ and surface temperature $T_s$ are external parameters. The atmospheric temperature profile is assumed to follow a dry adiabat until the stratospheric temperature $T_{strat}$ is reached. The latter was set to 200~K, close to the skin temperature for an early Earth with decreased solar flux and albedo of $\sim 0.3$. Sensitivity tests where a moist adiabat was used for the temperature profile showed only small differences in the column-integrated photolysis rates.

To solve (\ref{eq:photochem1}), a stiff ODE solver from the Fortran DLSODE package is used. For simplicity, the chemical network is restricted in this calculation to \ce{N} and \ce{H} species only. A total of 28 N-H reactions are included from various literature sources. Comprehensive calculations involving 4 or more chemical species (\ce{C}, \ce{N}, \ce{O}, \ce{H}) and 100s of chemical reactions for exoplanet applications indicate that \ce{N}-\ce{H} chemistry should dominate the photochemical conversion of \ce{NH3} to \ce{N2} \citep{Venot2012,Moses2013}. 

We neglect the possibility of ammonia shielding by an organic haze \citep{Sagan1997}. If present, this would allow higher \ce{NH3} buildup, 
but it would still require a substantial source of \ce{CH4}. 
UV shielding by \ce{CO2} was included in runs where the gas was present. 
In low temperature simulations, rainout of \ce{NH3} was included in the cool case via a parameterization based on present-day rainfall rates \citep{Kasting1982}. Rainout removes \ce{NH3} from the high atmosphere and hence decreases photolysis rates. 
Rayleigh scattering was included using standard parameterizations for \ce{N2} and \ce{H2} \citep{Pierrehumbert2011BOOK}. Because the aim was to derive a  lower limit on the total rates of photolysis, the actinic flux contribution was ignored here. This removed the need to perform a multiple scattering calculation. 

Photodissociation rates of \ce{NH3}, \ce{N2H4} and \ce{H2} were calculated using an XUV/UV solar spectrum  appropriate to 4.2~Ga, with data derived from a combination of direct solar observations and the Sun in Time project \citep{Ribas2005}. The solar zenith angle was set to 60$^\circ$ and a factor of $1/2$ applied to the incoming flux to account for diurnal averaging. {The model was validated by comparison with the results of previous calculations of \ce{NH3} photolysis for the Archean Earth \citep{Kuhn1979,Kasting1982}}.

{As an example of the model output, }
Figure~\ref{fig:photolysis_profs} shows steady-state molar concentrations of key species produced for an example case with 1~bar \ce{N2} atmosphere and 1\% \ce{H2}. As can be seen, \ce{NH3} concentrations decline up to $30$~km due to photolysis and rainout, although \ce{H2} photolysis causes $f_\ce{NH3}$ to increase in the uppermost regions of the atmosphere. 

Next, the steady-state ammonia photolysis loss rate ($\Phi_{loss,\ce{NH3}}$) was calculated as a function of the surface  $f_\ce{NH3}$  value, which was treated as a fixed parameter in the model. {To span the likely range of possibilities, two end-member atmospheric states are modeled: a hot, strongly reducing \ce{N2}-\ce{H2} atmosphere, and a cool, weakly reducing \ce{N2}-\ce{CO2} atmosphere. As can be seen from Figure~\ref{fig:photolysis_rates}, $\Phi_{loss,\ce{NH3}}$ ranges from just over $1\times10^{6}$ to close to $\Phi_{photo,max}$ and is most dependent on the \ce{NH3} surface molar concentration. The calculated minimum value for the hot \ce{H2}-rich case is somewhat lower than for the cool \ce{N2}-\ce{CO2} case due to \ce{NHx} radical reactions with \ce{H2}, which efficiently recycle photolysed \ce{NH3}.}

\subsection{Production of reduced nitrogen by thermolysis}

The final step is to calculate the range of surface temperatures and pressures for which thermochemical equilibrium is approached in the lower atmosphere. This was done by calculating the column-integrated rate of \ce{N2} reduction and comparing directly with the rate of ammonia photolysis. The thermochemistry of nitrogen reduction is still poorly constrained \citep{Moses2011,Venot2012}, so it was decided not to perform a fully coupled thermokinetic calculation here. Instead, we simply calculate the column-integrated rate of \ce{N2} reduction for fast and slow assumed rate-limiting reactions, and compare the results with the total photolysis rate for a range of surface pressures and temperatures. {The rate-limiting reactions used were}
\begin{equation}
\ce{N2 + H2 -> NH + NH}; \qquad  k_2(T) = 8.45\times10^{-8}e^{-81515.0/T}
\label{eq:slow_therm}
\end{equation}
and
\begin{equation}
\ce{N2 + H2 -> H2NN}; 
\qquad k_1(T) = 1.776\times10^{-16}T^{1.6}e^{-61580.0/T}
\label{eq:fast_therm}
\end{equation}
{for the slow and fast cases, respectively. Rate constants are in units of cm$^3$~molecule$^{-1}$~s$^{-1}$ in both cases. 
The rate for  (\ref{eq:slow_therm}) is based on experimental data and has been used to study the deep nitrogen cycle on Jupiter \citep{Prinn1981}, while that for (\ref{eq:fast_therm}) is derived from a fit to ab-initio calculations and has been used in state-of-the-art exoplanet atmospheric modeling \citep{Hwang2003,Moses2011}. }

The maximum column-integrated rate of \ce{NH3} production was assumed to be equal to twice the rate of \ce{N2} thermolysis 
\begin{equation}
\Phi_{\ce{NH3},thermo} =  2 \int^{z_{top}}_{0} n_\ce{N2}(z)n_\ce{H2}(z)\mathcal R[T(z)] \mbox{d}z
\label{eq:thermomakemax}
\end{equation}
where $n_\ce{N2}(z)$ is the number density of \ce{N2}, $n_{H_2}(z)$ is the number density of \ce{H2},  $\mathcal R[T(z)]$ is the rate of (\ref{eq:slow_therm}) or (\ref{eq:fast_therm}) 
as a function of temperature (see Supporting Information) and $z_{top}$ is the top of the model atmosphere. To calculate a maximum value, $n_\ce{H2}=n_\ce{N2}=0.5n_{tot}$ was assumed. {In a more realistic situation where the atmosphere was not dominated by equal parts \ce{H2} and \ce{N2}, the actual rate of \ce{NH3} production would be lower. } The rate of \ce{NH3} production in a  coupled model will also be lower than this if the initial reaction of \ce{N2} and \ce{H2} is not the rate-limiting step. In either case, this would increase the temperature and pressure required to achieve balance between surface thermochemical production and atmospheric photochemical destruction of reduced nitrogen. Equation (\ref{eq:thermomakemax}) gives a simple upper limit on the efficacy of \ce{N2} reduction by \ce{H2} as a function of atmospheric temperature and pressure. 

Fig.~\ref{fig:photo_thermo_bal} (left) shows contour plots of $\Phi_{\ce{NH3},thermo}$ in an \ce{H2}-rich atmosphere as a function of surface temperature and pressure. 
{Fig.~\ref{fig:photo_thermo_bal} (right) shows the surface temperature required for thermochemical production of \ce{NH3} to outpace its photochemical destruction in the atmospheric column, as a function of $p_{surf}$, for upper and lower limit cases. It can be assumed that above the limiting surface temperature, ammonia concentrations in the low atmosphere will be driven towards chemical equilibrium, which implies molar concentrations of 10-1000~ppm in this temperature range\footnote{Based on gas phase equilibrium calculations using the online NASA CEA database (\emph{http://cearun.grc.nasa.gov/}).}.  }

As can be seen, for all cases this requires surface temperatures in excess of $\sim$1000~K, with higher temperatures needed at lower pressures because of the dependence of chemical kinetics on gas density. {Even if a \ce{NH3} photolysis rate of only $10^4$~molecules/cm$^2$/s is assumed (the `extreme lower limit' case in Fig.~\ref{fig:photo_thermo_bal}) surface temperatures greater than 800~K are still required, thanks to the extremely high dependence of the rate constants of (\ref{eq:slow_therm}) and (\ref{eq:fast_therm}) on temperature.}

Typical solidus temperatures for Earth's mantle are $\sim$1500~K at 1~bar \citep{Elkins2012}. Hence {the key result of this analysis is that} conditions approaching those of a magma ocean at the surface are required for reduction of N if gas phase chemistry is the limiting factor. 
{Magma ocean surface conditions on Earth occurred during and just after formation \citep{Matsui1986,Zahnle2010}. }
The moon-forming impact transformed much of Earth's surface into a molten state \citep{Stewart2014}.  The integrated effect of smaller impacts is also sufficient to increase surface temperatures above 1500~K if the thermal blanketing of a \ce{H2O} vapour (or \ce{H2}) atmosphere is taken into account. Hydrogen absorbs strongly in the infrared above pressures of $\sim0.2$~bar due to \ce{H2}-\ce{H2} and  \ce{N2}-\ce{H2} collision-induced absorption, so a thick enough \ce{H2} envelope would provide extremely effective greenhouse warming\footnote{An approximate estimate of the equilibrium surface temperature for an \ce{H2}-dominated atmosphere can be given by  $T_s = T_e(p_s/p_e)^{R/c_p} $ where $R$ is the specific gas constant, $c_p$ is the specific heat capacity at constant volume, $T_e$ is the emission temperature, $p_e$ is the emission pressure (around 0.2~bar for \ce{H2}) and $p_s$ is surface pressure. Taking $T_e$ to be the equilibrium temperature given albedo $A=0.3$ and a faint young Sun ($T_e=235$~K),  and $R/c_p=0.289$, we get $T_s = 1000$~K for  around 30~bar of \ce{H2}. Detailed radiative-convective calculations (results not shown) indicate that when \ce{H2O} is included the required pressure is around 100~bars of \ce{H2} for planetary albedos in the range 0.3-0.7, due to the effect of moist convection on the adiabatic lapse rate.}.

Under magma ocean conditions, reduced atmospheric nitrogen can be transported to the mantle rapidly, because the high Rayleigh numbers typical to magma ocean conditions  generally ensure rapid mixing \citep{Solomatov2000}. In addition, previous experimental work has indicated that nitrogen solubility is greatly increased in reducing magmas and upper mantle minerals \citep{Libourel2003,Kadik2011,Li2013}.  {At relatively high oxygen fugacities (QFM and greater), \ce{N2} is the dominant form of N in the melt, and it behaves similarly to argon, with a solubility of around 0.1~ppmw at 1~bar \ce{N2} pressure \citep{Libourel2003}.  As oxygen fugacity decreases to the IW buffer and lower, chemical bonding of N to the silicate melt becomes important and solubility increases by orders of magnitude. Earth's mantle \ce{fO2} has been close to the QFM buffer since around 3.8~Ga \citep{Delano2001}, and perhaps as far back as 4.35~Ga  \citep{Trail2011}, but immediately after formation the mantle would have been highly reducing, at the IW buffer or lower \citep{Wade2005}.}  
The idea that dissolution of N in an early reducing magma ocean led to significant incorporation of N in the mantle has been proposed previously [e.g., \citep{Marty2012,Li2013}]. The  simplicity and robustness of this mechanism compared to the others we have analyzed here lead to the conclusion that it is the most plausible explanation for early incorporation of N into the mantle.

\subsection{Catalytic effects and \ce{CO2} atmospheres}
We have only considered pure N-H chemistry, but catalytic effects could also have been important to Hadean N fixation. Notably, in the Haber process, nitrogen is converted to ammonia in the presence of metallic iron at around 200~bar and $700$~K \citep{Erisman2008}. A similar process could have occurred in the Hadean, weakening the surface temperature and pressure requirements. However, the Haber process requires the presence of metallic iron, which in turn implies a crustal oxygen fugacity around the iron-w\"ustite (IW) buffer or lower. Hence it does not change the basic requirement of a highly reducing surface and atmosphere for abiotic N drawdown. 

Similar considerations apply to other catalytic processes such as lightning-driven thermochemistry. Using modern-day rates of energy dissipation by lightning, \cite{Chameides1981} estimated a peak \ce{HCN} production rate in the Archean of around  
6$\times10^8$~molecules/cm$^2$/s 
given an atmosphere with  $5\%$ \ce{H2} and C:O ratio greater than 1 (which is extremely reducing). This could lead to drawdown of around $0.6$~bar of \ce{N2} in 1~Gy, although the conversion rate of atmospheric HCN to fixed nitrogen in sediment would need to be close to 100\% for it to work.

One other possible scenario for the Hadean immediately after the Moon-forming impact  involves a thick  \ce{CO2} atmosphere (10-100~bar). Under such an atmosphere, a magma ocean could persist for around 10~My, after which time surface temperatures and atmospheric \ce{CO2} levels would likely decline rapidly \citep{Sleep2001,Zahnle2010}. A  \ce{CO2}-dominated atmosphere could shield ammonia from UV photolysis effectively and would cause enhanced ammonia rainout rates by decreasing raindrop pH. It would also decrease ocean pH levels, probably to the extent where most ammonia in the atmosphere-ocean system would be converted to aqueous \ce{NH4^+}. 
For rapid N drawdown to occur in this scenario, however, an efficient abiotic means of reducing \ce{N2} in the first place is still required.

\section{Nitrogen equilibration on Venus: The water loss redox pump }\label{sec:Opump}

If a Hadean abiotic process is the explanation for Earth's atmosphere-mantle nitrogen partitioning, the above analysis should also apply to other planets. The case of Venus is particularly interesting, because Venus is similar in size to Earth, but it possesses around 3.4 times as much nitrogen in its atmosphere scaled by mass \citep{Hoffman1980,Goldblatt2009}. What could explain the difference?

The most obvious difference between Earth and Venus is their distance to the Sun. Because it receives a greater solar flux, Venus most likely passed through a runaway greenhouse phase, either immediately after its formation \citep{Gillmann2009,Hamano2013} or later on, once the solar luminosity increased past a critical threshold \citep{Ingersoll1969}. In either case, extensive water loss accompanied by buildup of oxygen in the atmosphere would have been the result. The abundance of \ce{O2} in the Venusian atmosphere today is $<3$~ppm \citep{Mills1999}, so the vast majority of this oxygen must have been absorbed by the crust and mantle.

The redox state of the Venusian crust is close to the magnetite-hematite boundary 
based on spectroscopic observations of surface minerals by the Venera spacecraft \citep{Florensky1983,Fegley1997}. Because Venus underwent a major resurfacing event ca. 0.3~Ga \citep{Strom1994} and hydrogen loss rates in the present era are low ($\sim 1\times10^{26}$~atoms/s, or $1.0\times10^{-5}$ terrestrial oceans in the last 0.3~Gy) \citep{Lammer2006}, the Venusian surface redox state should be close to that of its mantle. In contrast, the oxygen fugacity of Earth's upper mantle is closer to the more reducing QFM buffer. 

Based on these observations, we can propose the following explanation for the difference in Venus and Earth's atmospheric nitrogen inventories. Both planets have highly reducing steam atmospheres (state A in Figure~\ref{fig:redox_pump}) during and shortly after formation. Earth cools relatively rapidly and forms liquid oceans, passing to state C: a planet with around 1~bar of atmospheric \ce{N2}, smaller amounts of atmospheric \ce{H2}, and a moderately oxidising mantle. In the terminology of \cite{Hamano2013}, it is a Type I planet. In contrast, Venus undergoes a sustained runaway greenhouse phase (state B), leading to extensive photolysis of \ce{H2O} followed by loss of hydrogen to space. The oxygen liberated from \ce{H2O} photolysis is then absorbed by the mantle, raising its oxygen fugacity and expelling nitrogen into the atmosphere as a result. The end state is a dry planet with a greater atmospheric \ce{N2} inventory (state D).

{An estimate of the total oxidation of Venus during an early period of \ce{H2O} photolysis can be made by noting that in the Sun's saturated phase of XUV emission, which probably lasted around 100~My \citep{Ribas2005}, XUV levels would have been high enough to drag oxygen along with the escaping hydrogen \citep{Zahnle1986b,Chassefiere1996}. In this escape regime, the rate of oxygen buildup is approximately determined by the diffusion of O atoms downwards from the upper atmosphere \citep[e.g., ][]{Luger2015}, with the flux given by
\begin{equation}
\Phi_\ce{O}\approx \frac{5 b m_\ce{H}g}{k_BT}. 
\end{equation}
Here $m_\ce{H}$ is the mass of one hydrogen atom, $g$ is Venusian gravity in the upper atmosphere, $k_B$ is Boltzmann's constant, $T$ is the temperature of the atmosphere in the region of peak diffusion and $b=b(T)$ is the binary diffusion coefficient for O in H  \citep{Zahnle1986}. Assuming $T\approx 300-600$~K, $\Phi_\ce{O}\approx 3.8-7.6\times10^{12}$~molecules/cm$^2$/s. Hence 1.2-2.4 terrestrial oceans (TO) worth of oxygen could  have been liberated from \ce{H2O} on Venus during the first 100~My. For comparison, Earth's estimated total \ce{H2O} inventory is of order 1-4~TO \citep{Hirschmann2006}.}

How much liberated O is necessary to change the oxidation state of the mantle? If Venus' early upper mantle was 10\% of the total planetary mass and contained $5$~wt\% iron in the form of \ce{FeO}, oxidation of all of the \ce{FeO} to hematite (\ce{2FeO + 0.5 O_2 \to Fe_2O_3})
would require 2.8~TO, given that 1~TO is $7.8\times10^{22}$~moles~O. This is a lot of oxygen. 
However, the amount of water loss required to affect early nitrogen partitioning will be lower than this value. First, just after formation the redox state of rocky planet mantles is below even the IW buffer \citep{Wade2005}. 
In the absence of buffering by an abundant element such as \ce{Fe}, much lower levels of O buildup will then lead to significant mantle $f_\ce{O2}$ changes. Differences of 1-2 in log$_{10}f_\ce{O2}$ below the IW buffer cause variations in N solubility of 0.1-10~ppm at 1~bar \citep{Libourel2003}. In early highly reducing conditions, therefore, the oxidation of Venus due to water loss would have a much greater effect than later on. Second, planetary interiors are highly heterogeneous. If water loss on early Venus led to mixing of highly oxidised surface material with reduced, N-rich material in the deep interior, this could also cause increased outgassing of N over time even if the oxidation state of the upper mantle remained constant. Increased oxidation via water loss hence appears easily capable of explaining the 3.4$\times$ greater atmospheric inventory of Venus compared to the Earth. 

Alternative explanations for the differing atmospheric N inventories of Earth and Venus involving delivery or escape processes are possible, but less compelling. Venus has atmospheric \ce{Ne} and \ce{^{36}Ar} abundances several tens of times greater than Earth, suggesting that it captured a larger component of the nebula during formation \citep{Wieler2002}. However, this nebular addition cannot have increased Venus' total atmospheric nitrogen inventory significantly. There is approximately $3.2\times10^{17}$~mol of \ce{^{36}Ar} in the Venusian atmosphere vs. $3.7\times10^{20}$~mol of \ce{N2} \citep{Hoffman1980,Wieler2002,Lodders2003}. 
The \ce{N}/\ce{^{36}Ar} molar ratio in the solar photosphere is around 22.4,  
which implies that a maximum of about 1.0\% of Venus's atmospheric nitrogen could have a nebular origin. This is consistent with Venus' essentially chondritic \ce{^{15}N}/\ce{^{14}N} ratio, which suggests the bulk of its nitrogen was supplied from the same source as Earth's \citep{Hunten1983}.

{Regarding loss processes, Venus' closer proximity to the Sun means that its early XUV-driven escape rates were greater than Earth's, both for hydrogen and potentially for N {via hydrodynamic drag} \citep{Chassefiere1996}. Earth underwent the Moon-forming impact, which could have removed a substantial portion of the atmosphere, depending on factors such as the initial angular momentum of the protoearth and presence/absence of an ocean \citep{Genda2005,Cuk2012}. However, Venus must also have lost significant amounts of its atmosphere to space by impact erosion during the late stages of its formation. In addition, impact-driven loss of \ce{N2} to space from Earth after the Moon-forming impact would cause loss of noble gases {such as \ce{^{36}Ar}} by the same proportion. However {as mentioned in the Introduction}, Earth is depleted in atmospheric N vs. noble gases relative to chondritic material. This suggests extensive loss of N to space on Earth as an explanation for its lower atmospheric N inventory vs. Venus is also problematic, unless Earth's noble gas inventory is dominated by a late cometary contribution \citep{Marty2012}.}

\section{Conclusion}\label{sec:conc}

Taken together, the existing geological constraints on Earth's nitrogen present a paradox: large amounts of N appear to be present in the mantle, there is a high present-day flux to the mantle from the surface, but the amount of \ce{N2} in the atmosphere has not changed significantly since 3-3.5~Ga. There are several possible resolutions to this problem. The first is that some of the observations are incorrect. In future, study of a wider range of proxies for paleopressure over multiple eras and more detailed investigation of heterogeneity in present-day fluxes and reservoirs will be necessary to eliminate this possibility.

Starting from the assumption that the latest constraints on paleopressure and mantle N inventory are correct, we have considered a range of possible mechanisms for nitrogen drawdown in Earth's early history. It has been shown that for multiple reasons, biologically driven subduction of bars of nitrogen in the early Archean is unlikely. Instead, by far the most effective way to fix atmospheric N rapidly is via direct reduction in a hot Hadean atmosphere. {Hence if $p_\ce{N2}$ was indeed low 3-3.5~Ga, Earth's mantle N must have been emplaced early, when conditions were extremely hot and reducing. The simplest way for this to occur is via direct diffusion into a magma ocean, when atmospheric-interior exchange rates would have been rapid. A new redox pump mechanism has been proposed that can naturally explain the differences in atmospheric N inventory between Venus and Earth via enhanced mantle oxidation following water loss in this scenario.}

Interestingly, Mars has an extremely low atmospheric \ce{N} inventory ($\sim 5\times10^{-8}$~wt\% of the total mass, vs. $6.7\times10^{-5}$~wt\% for Earth)  \citep{Mahaffy2013}. Escape to space goes some way towards explaining the difference, but may not account for all of it \citep{Jakosky1994}. The topmost region of Mars' surface is highly oxidized, but Mars' low received solar flux prohibits a long-lived early steam atmosphere after formation. Indeed, the oxygen fugacity of the Martian mantle is even lower than that of Earth \citep{Wadhwa2001}, possibly as a result of its smaller size and rapid core formation \citep{Wade2005}. Storage in the mantle may therefore also help explain Mars' nitrogen-poor atmosphere.

The results described here have several important wider implications. The redox pump explanation for Venus' nitrogen inventory has implications for predicting the evolution of other volatiles (e.g. carbon) both on Venus and on close-in, low mass exoplanets such as Gliese 1132b \citep{BertaThompson2015}. It is also testable, in principle, by future \emph{in situ} measurements of the N inventory of the Venusian crust. In addition, if an early reducing atmosphere is necessary to explain Earth's nitrogen budget, this has implications for prebiotic chemistry and the emergence of life. Future theoretical studies should investigate these implications in more detail.

\section{Acknowledgments}
The author thanks Eric Hebrard for his advice and sharing of reaction rate data during construction of the photochemical code and the anonymous reviewers for their comments, which greatly improved the quality of the paper. This manuscript has benefited from conversations with many researchers, including N. Dauphas, A. Knoll, H. Lammer, B. Marty, A. Pearson, D. Schrag and K. Zahnle. 


\begin{figure}[h]
	\begin{center}
		{\includegraphics[width=3.0in]{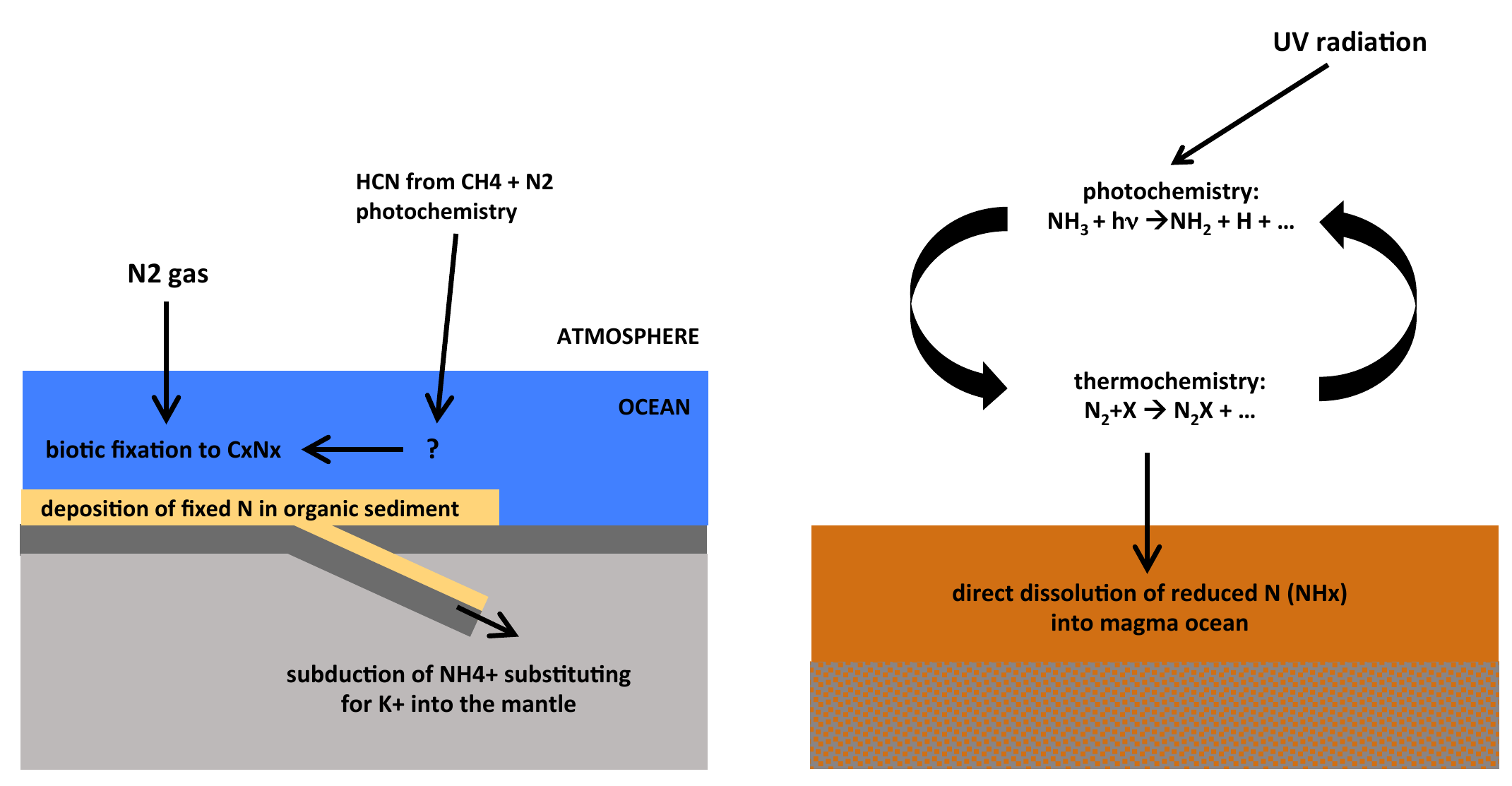}}
	\end{center}
	\caption{(left) Biological and (right) abiotic mechanisms capable of transferring atmospheric nitrogen to early Earth's mantle. Biological mechanisms are possible from the origin of nitrogen fixation and methanogenesis metabolisms onwards, while the abiotic mechanism shown would have operated in the early Hadean, during and just after accretion.	}
	\label{fig:photo_schem}
\end{figure}

\begin{figure}[h]
	\begin{center}
		{\includegraphics[width=3.0in]{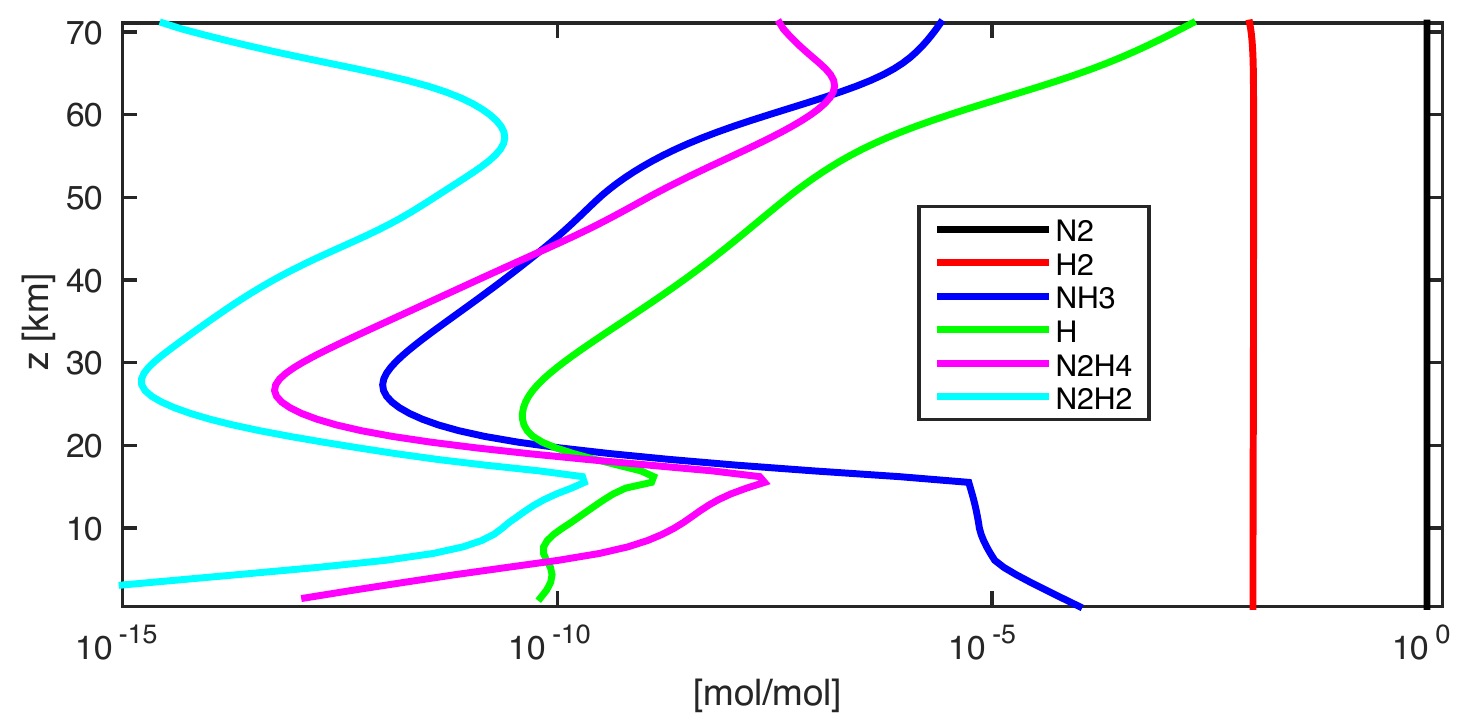}}
	\end{center}
	\caption{Example results from the photochemical model. Steady state atmospheric molar concentrations of major species for a \ce{N2}-dominated atmosphere with $T_s=300$~K, $p_s=1$~bar, $10^4$~ppmv \ce{H2}, surface \ce{NH3} concentration of $100$~ppmv, and eddy diffusion and rainout included based on present-day parametrizations.}
	\label{fig:photolysis_profs}
\end{figure}

\begin{figure}[h!]
	\begin{center}
		{\includegraphics[width=3.0in]{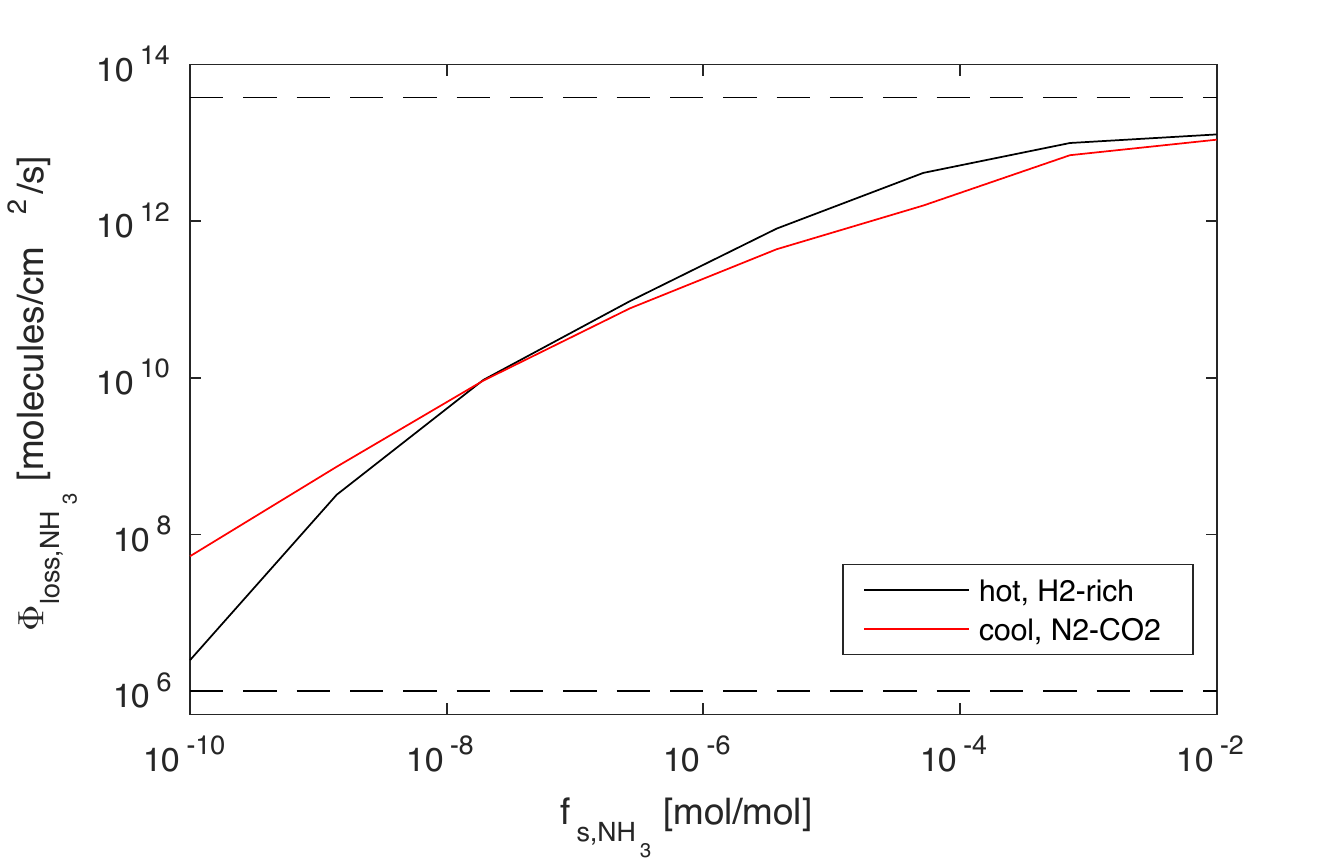}}
	\end{center}
	\caption{Predicted ammonia destruction rates as a function of surface \ce{NH3} molar concentration. The black and red lines  show results from the photochemical code for a hot hydrogen-rich atmosphere ($f_\ce{H2}=0.5$~mol/mol, $f_\ce{N2}=0.5$~mol/mol, $T_s=700$~K, $p_s=30$~bar) with constant eddy diffusion $\kappa_D = 10^5$~cm$^2$~s$^{-1}$ and a cool \ce{N2}-\ce{CO2} atmosphere ($f_\ce{N2}=0.9$~mol/mol, $f_\ce{CO2}=0.1$~mol/mol, $T_s=300$~K, $p_s=1$~bar) with present-day eddy diffusion profile, respectively. The dashed lines show the  upper and lower limits for photolysis rates derived from the results.}
	\label{fig:photolysis_rates}
\end{figure}

\begin{figure}[h!]
	\begin{center}
		{\includegraphics[width=3.0in]{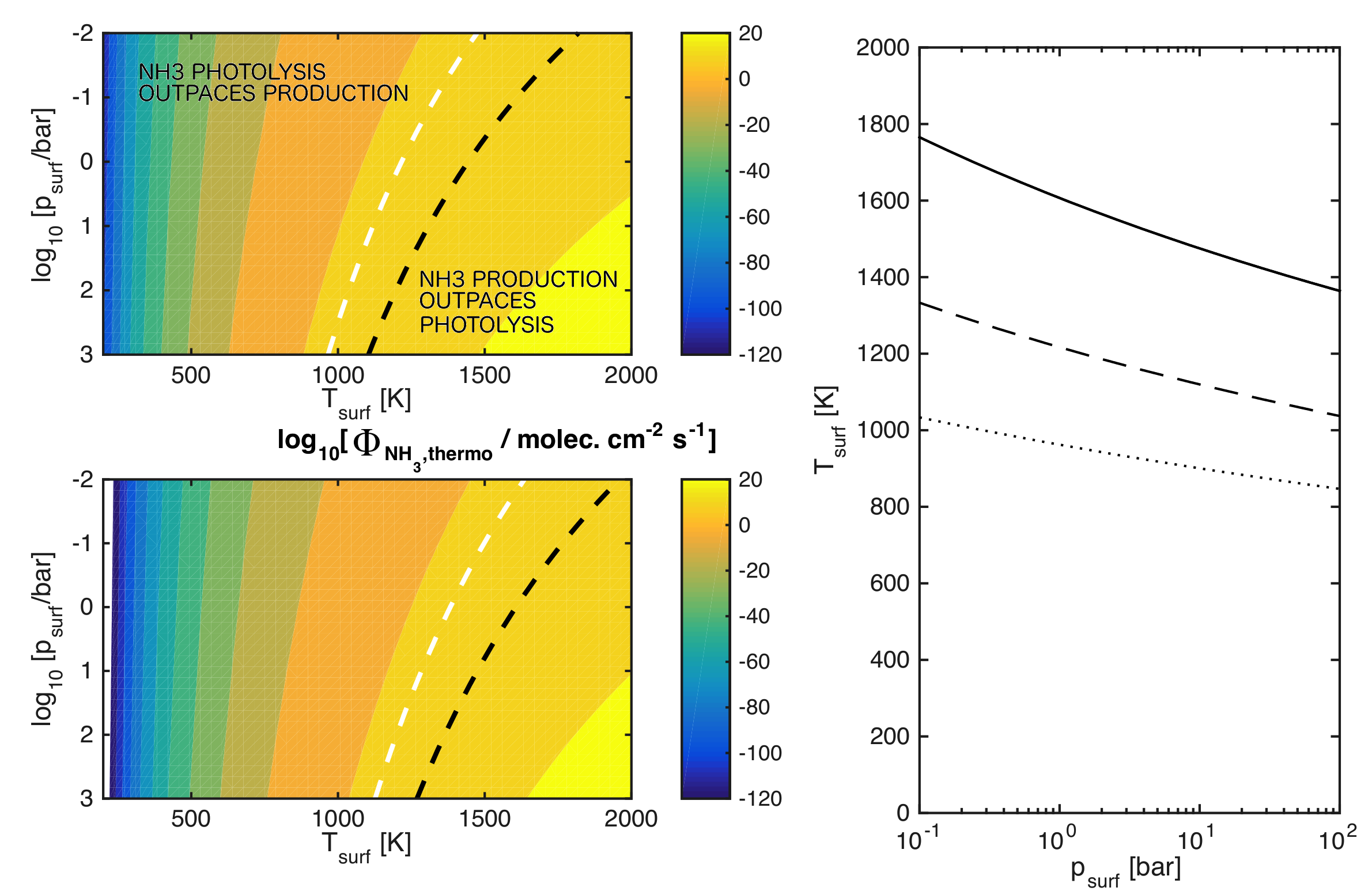}}
	\end{center}
	\caption{(left) \ce{NH3} column production rate via atmospheric thermochemistry as a function of  surface temperature $T_s$ and surface pressure $p_s$, assuming that (top) equation (\ref{eq:slow_therm}), (bottom) equation (\ref{eq:fast_therm}) in is the rate-limiting step. The black and white lines show the $T_{s}$-$p_{s}$ curve for which this estimated thermolysis rate equals the maximum and minimum column-integrated \ce{NH3} photolysis rate from the photochemical model, respectively (see Fig.~\ref{fig:photolysis_rates}). {(right) Surface temperature required for ammonia production to outpace photolysis. The solid line is the upper limit assuming slow \ce{N2} thermolysis [reaction (5)] and fast \ce{NH3} photolysis ($\Phi_{loss,\ce{NH3}}=\Phi_{photo,max}$) and the dashed line assumes fast \ce{N2} thermolysis [reaction (6)] and slow \ce{NH3} photolysis ($\Phi_{loss,\ce{NH3}}=\Phi_{photo,min}$). The dotted line is an extreme lower limit with fast \ce{N2} thermolysis and  $\Phi_{loss,\ce{NH3}}=\Phi_{photo,min}/10^2$.}} 
	\label{fig:photo_thermo_bal}
\end{figure}

\begin{figure}[h]
	\begin{center}
		{\includegraphics[width=3.0in]{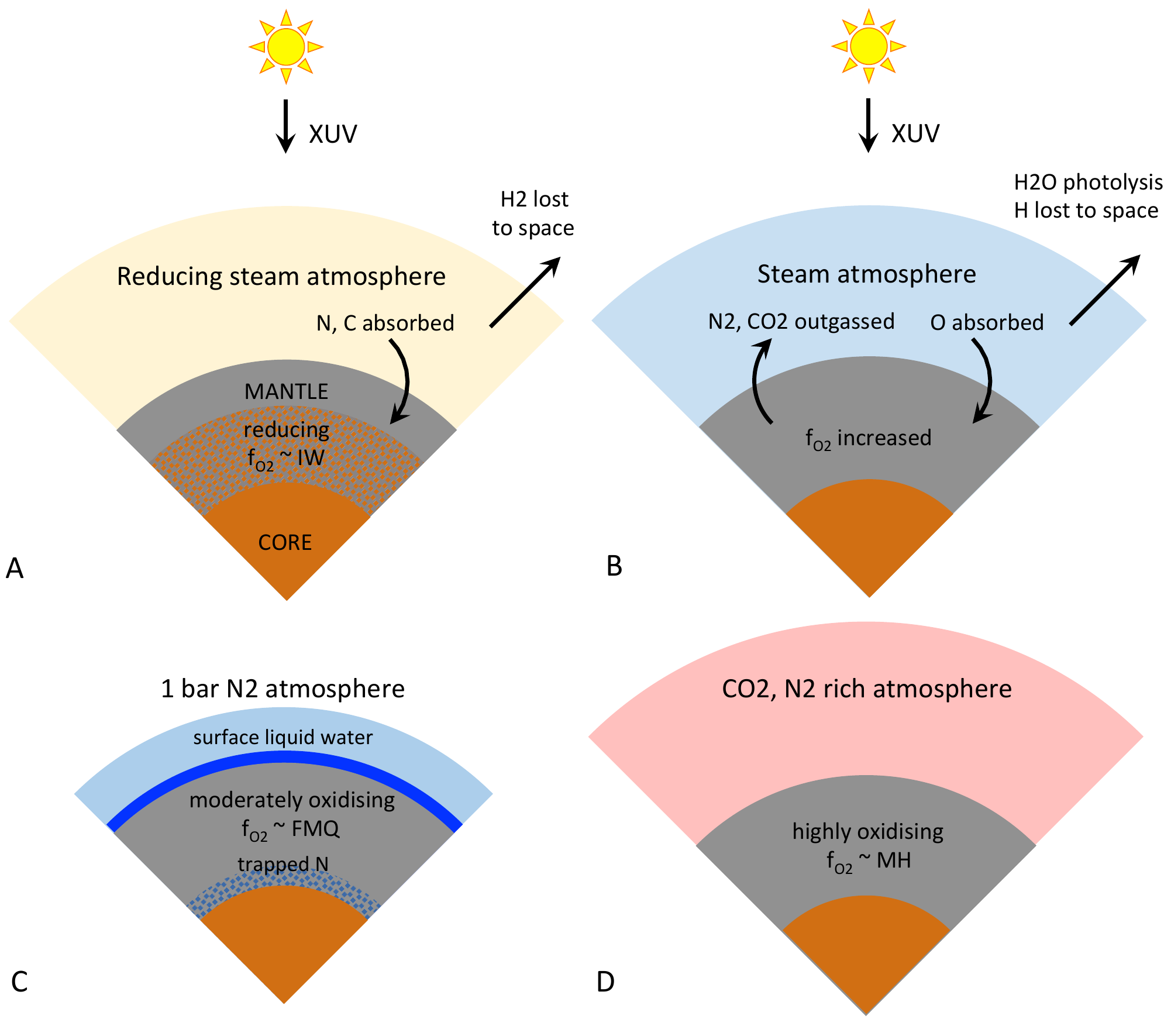}}
	\end{center}
	\caption{Schematic of the water loss redox pump explanation for the differing atmospheric \ce{N2} inventories of Earth and Venus. Both planets start in state A. Earth evolves directly to state C, whereas Venus passes through state B before ending up in D. See text for full description.}
	\label{fig:redox_pump}
\end{figure}

\end{document}